\documentclass[fleqn]{2017SCGE}
\newcommand\aap{A\&A}                
\newcommand\apj{ApJ}                 
\newcommand\apjl{ApJ}                
\newcommand\mnras{MNRAS}             
\newcommand\nat{Nature}              
\newcommand\prc{Phys. Rev.~C}        
\newcommand\prd{Phys. Rev.~D}        

\begin{document}

\ensubject{subject}

\ArticleType{Article}
\Year{2018}
\Month{February}
\Vol{60}
\No{1}
\DOI{xxx}
\ArtNo{000000}
\ReceiveDate{February xx, 2018}
\AcceptDate{xx xx, 2018}

\title{Causal propagation of signal in strangeon matter}

\author[1,2]{Jiguang LU}{lujig@nao.cas.cn}%
\author[3]{Enping ZHOU}{}
\author[4,5]{Xiaoyu LAI}{}
\author[3,6]{Renxin XU}{}

\AuthorMark{Lu J G}

\AuthorCitation{Lu J G, Zhou E P, Lai X Y and Xu R X}

\address[1]{National Astronomical Observatories, Chinese Academy of Sciences, Beijing 100012, China}
\address[2]{Key Laboratory of Radio Astronomy, Chinese Academy of Science, Beijing 100012, China}
\address[3]{School of Physics and State Key Laboratory of Nuclear Science and Technology, Peking University, Beijing 100871, China}
\address[4]{School of Physics and Mechanical \& Electrical Engineering, Hubei University of Education, Wuhan 430205, China}
\address[5]{Xinjiang Astronomical Observatory, Chinese Academy of Sciences, Urumqi, Xinjiang 830011, China}
\address[6]{Kavil Institute for Astronomy and Astrophysics, Peking University, Beijing 100871, China}


\abstract{%
The state equation of strangeon matter is very stiff due to the non-relativistic nature of and the repulsing interaction between the particles, and pulsar masses as high as $\sim 3M_\odot$ would be expected.
However, an adiabatic sound speed, $c_s=\sqrt{\partial P/\partial \rho}$, is usually superluminal for strangeon matter, and dynamic response of strangeon star (e.g., binary merger) could not be tractable in numerical simulations.
We examine signal propagation in strangeon matter, and calculate the propagation speed, $c_{\rm signal}$, in reality.
It is found that as the causality condition is satisfied, $c_{\rm signal}<c$, and the signal speed as a function of stellar radius is presented.
}



\maketitle


\begin{multicols}{2}
\section{Introduction}\label{sect:intro}

The nature of pulsar depends on the understanding of the state of supranuclear matter, to be related to the non-perturbative behavior of fundamental strong interaction, and remains unclear even it is more than a half century after the discovery.
Nevertheless, this unknown state could soon be understood in the era of gravitational wave astronomy, especially after the event of GW170817 which is caused by the merger of
binary compact stars~\cite{abbo17}.
This kind of events can surely help in distinguishing the equation
of state (EoS) models of compact star (see~\cite{baio17} for a review).
In fact, the post-merger gravitational wave signal and the electromagnetic counterparts
are closely related to the shocks and ejected material accompanying binary compact star mergers,
while the
\Authorfootnote

\noindent dynamical response there depends on the sound speed in compact star matter~\cite{rezz13}.
Of course, the sound speed depends on the EoS of compact star matter.
Therefore, it is essential to have a proper calculation of the sound speed in order
to study binary compact star mergers with numerical relativity.

There are many speculations about the state equation of cold supranuclear-density matter, due to
lack of well understanding of quantum chromodynamics at low energy scale, and socalled strange matter conjecture is among them.
It says that 3-flavored strange matter (composed by {\em free} $u$, $d$ and $s$ quarks)
could even more stable than 2-flavored nucleon matter~\cite{bodm71,witt84}, based on which various models of 3-flavored strange star were put forward.
The MIT bag model was a widely used, treating the quarks free and relativistic,
and this model was always adopted to describe the matter state of pulsar~\cite{alco86,haen86}.
In this case, the maximum mass of pulsar can hardly reach $2~M_{\odot}$.
Then, the observation of the $1.97~M_{\odot}$ pulsar J0715$+$1807~\cite{demo10} and
$2.01~M_{\odot}$ pulsar J0348$+$0432~\cite{anto13} marginally ruled out the quark star model.
In fact, even in neutron star model, $2~M_{\odot}$ is still too large to approach
due to the ``hyperon puzzle''~\cite{scha08,Bombaci17}.
Nonetheless, another strange matter model in which the quarks are bound in clusters
(which are formerly known as strange quark-clusters~\cite{xu03},
just like nucleons but with strange quarks,
hereafter, strangeons) allows the maximum mass of pulsar to be much larger~\cite{lai09}.
Therefore, the strangeon star model is more favorable in comprehending the mass of the pulsars
than other models.
Besides, the strangeon star model can also solve some problems which are
difficult to understand in other models, e.g., bi-drifting of sub-pulses~\cite{qiao04},
non-atomic feature in spectrum of X-ray dim isolated neutron stars (XDINSs,~\cite{lu13}),
two types of glitches in normal pulsar and AXP/SGR~\cite{zhou14}
and the optical/UV excess of XDINSs~\cite{wang17}.

However, the strangeon star model is sometimes repelled because its matter state is too stiff.
Based on the conventional formula, the sound speed $c_{\mathrm{s}}=\sqrt{\partial P/\partial\rho}$ in such stiff matter would even exceed the speed of light~\cite{blud68,blud70,bayi80,keis96}, and the maximum mass of a compact star would then never larger than $3.2~M_{\odot}$~\cite{rhoa74}.
Obviously, the maximum mass of a strangeon star conflicts to above result~\cite{lai09,lai13,guo14}.
In fact,  Caporaso \& Brecher pointed out that it is possible to construct lattice model with $\partial p/\partial\rho>c^2$ and a subluminal signal speed \cite{capo79}.
However, that work was under the assumption of electromagnetic interaction and
didn't give the expression of signal speed.
Here we want to put forward a signal speed in a more general case, which at least
can be correctly adopted in strangeon matter and is truly necessary in simulating strangeon star merger~\cite{lai17} with numerical relativity.

A strangeon is much more massive than a nucleon. Therefore, in matter with similar
mass density, the quantum wave packet of a particle in strangeon matter is smaller
than that in nucleon matter.
Thus, the strangeon could be regarded as a classical particle localized
rather than a quantum wave packet.
Then, in this paper, for the sake of simplicity, we will consider the oscillation propagation
in a 1-D discrete chain.
In \S~\ref{sect:calculation}, the sound speed in
a 1-D chain of particles is derived theoretically, and some discussions about the speed
are listed in \S~\ref{sect:discussion}.
The results are summarized in \S~\ref{sect:summary}.

\section{A model to calculate signal speed}
\label{sect:calculation}

Consider a 1-D chain of particles, and along the chain the particles
can vibrate slightly.
Then the position of the $n$-th particle can be expressed as the function of time $t$,
\begin{equation}
x_n=f(n,t)+l(n),
\end{equation}
where $l(n)$ is the average position of the $n$-th particle, $f(n,t)$ is its
relative displacement which is zero-mean.
Assume a two-body short-range (which means that the interaction only acts on nearby
particles) repulsive conservative interaction $F(x)$, which is only related to the distance between two particles $x$.
Then the force on the $n$-th particle caused by the posterior or the prior particle
can be expressed,
\begin{equation}
F_{n+}=F(\left|x_{n+1,~\mathrm{ret}}-x_n\right|),~
F_{n-}=F(|x_{n-1,~\mathrm{ret}}-x_n|),
\end{equation}
where the subscript ``ret'' means that this force is a retarding force because of the
propagation of the force field.
The propagation speed of the force field is assumed to be the speed of light $c$ in this paper.
With the assumption of small amplitude, $f(n,t),~f(n-1,t)\ll l(n)-l(n-1)$,
we can expand the force to the first order of the distance between particles at the average position.

\begin{equation}
\begin{split}
F_{n+}&=F\Bigg(f\left(n+1,t-\frac{l(n+1)-l(n)}{c}\right)-f(n,t)\\
&~~~~~~~~~~~~~~~~~~~~~~~~~~~~~~+[l(n+1)-l(n)]\Bigg)\\
&\approx F(l(n+1)-l(n))+\Bigg[f\left(n+1,t-\frac{l(n+1)-l(n)}{c}\right)\\
&~~~~~~~~~~~~~~~~~~~~~~~~~~~~~~-f(n,t)\Bigg]\frac{\partial F}{\partial x}\bigg|_{x=l(n+1)-l(n)},~
\end{split}
\end{equation}
\begin{equation}
\begin{split}
F_{n-}&=F\Bigg(f(n,t)-f\left(n-1,t-\frac{l(n)-l(n-1)}{c}\right)\\
&~~~~~~~~~~~~~~~~~~~~~~~~~~~~~~+[l(n)-l(n-1)]\Bigg)\\
&\approx F(l(n)-l(n-1))+\Bigg[f(n,t)\\
&-f\left(n-1,t-\frac{l(n)-l(n-1)}{c}\right)\Bigg]\frac{\partial F}{\partial x}\bigg|_{x=l(n)-l(n-1)}.
\end{split}
\end{equation}
On the other hand, the resultant force on the $n$-th particle can be expressed as following,
$$
F_n=\frac{\mathrm{d}^2f}{\mathrm{d}t^2}=F_{n-}-F_{n+}\approx F(l(n)-l(n-1))-F(l(n+1)-l(n))
$$
$$
+\left[f(n,t)-f\left(n-1,t-\frac{l(n)-l(n-1)}{c}\right)\right]\frac{\partial F}{\partial x}\bigg|_{x=l(n)-l(n-1)}
$$
\begin{equation}
-\left[f\left(n+1,t-\frac{l(n+1)-l(n)}{c}\right)-f(n,t)\right]\frac{\partial F}{\partial x}\bigg|_{x=l(n+1)-l(n)},
\label{eqdyn}
\end{equation}
where $m$ is the mass of each particle.
Taking the average of Eq.~\ref{eqdyn}, we can get $F(l(n)-l(n-1))-F(l(n+1)-l(n))=0$.
The repulsive interaction is generally monotonous versus the distance $x$ (in a small range),
which means that $l(n)-l(n-1)=l(n+1)-l(n)$, i.e., the interparticle spacing is regular.
This constant is marked as $l$ below.

To Calculate the sound speed, firstly, the wave propagation process in frequency domain is considered similar to traditional method.

\subsection{Frequency Domain Oscillation Propagation in Infinite Chain}
\label{Oscillation}

In the frequency domain, the Eq.~\ref{eqdyn} can be presented as
\begin{equation}
\frac{g(n+1,\omega)+g(n-1,\omega)}{g(n,\omega)}=2\left(1-\frac{m\omega^2}{2\frac{\partial F}{\partial x}\big|_{x=l}}\right)\exp\left(\mathrm{i}\frac{l\omega}{c}\right),
\label{eqfreq}
\end{equation}
where $g(n,\omega)$ is the complex amplitude of the $n$-th particle at frequency $\omega$, and $\frac{\partial F}{\partial x}\big|_{x=l}\approx -3(1-2\nu)\frac{m}{l^2}\left(\frac{\partial P}{\partial\rho}\right)_T$, $P$ and $\rho$ are the internal pressure and density of the chain respectively, $\nu$ is the Poisson's ratio, and the subscript $T$ means that the derivative is taken isothermally.
Considering that Poisson's ratio for a perfectly isotropic elastic material is 0.25 and the adiabatic index
of 1-D matter here is 3 (degree of freedom is 1), the $\frac{\partial F}{\partial x}\big|_{x=l}$ could be
expressed as $-\frac{m}{2l^2}\left(\frac{\partial P}{\partial\rho}\right)_S$, the subscript $S$ means that the derivative is taken isentropically.
The stable solution of Eq.~\ref{eqfreq} is (the diverge branch is abandoned)
\begin{equation}
\frac{g(n+1,\omega)}{g(n,\omega)}=b-\sqrt{b^2-1},
\label{eqratio}
\end{equation}
where $b=\left[1-\frac{l^2\omega^2}{\left(\frac{\partial P}{\partial\rho}\right)_S}\right]\exp\left(\mathrm{i}\frac{l\omega}{c}\right)$.
With the phase variation shown in above equation,
the apparent phase velocity and group velocity of the oscillation can be calculated as following,
\begin{equation}
c_{\mathrm{p}}=\frac{l\omega}{-\arg\left[\frac{g(n+1,\omega)}{g(n,\omega)}\right]}
\label{eqcsp}
\end{equation}

\begin{equation}
c_{\mathrm{g}}=\frac{c_{\mathrm{p}}^2}{c_{\mathrm{p}}-\omega\frac{\mathrm{d}c_{\mathrm{p}}}
{\mathrm{d}\omega}}.
\end{equation}

As shown in Eq.~\ref{eqratio}, the amplitude of the oscillation damps while propagating, and this implicates that there is reflection wave in the chain. In this case neither phase velocity nor group velocity represents the velocity of the signal propagation. Hence we should calculate the signal propagation in time domain.

\subsection{Time Domain Impulse Response in Finite Chain}
\label{response}

\begin{figure*}
  \centering
   \includegraphics[width=0.9\linewidth]{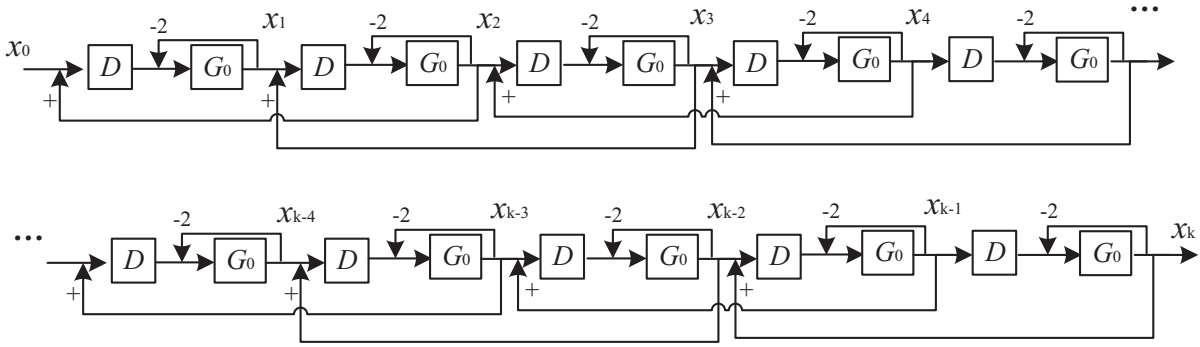}
   \caption{{\small Block diagram of signal propagation. Each node is the displacement of each particle in the chain, $D$ is the delay element and $G_0$ is the proportional derivative element.}}
   \label{fig1}
\end{figure*}

\begin{figure*}
  \centering
   \includegraphics[width=0.9\linewidth]{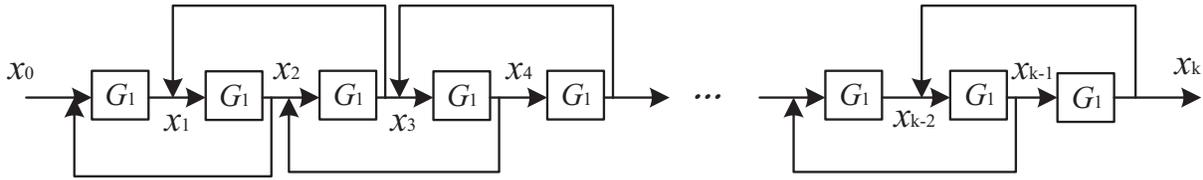}
   \caption{{\small Simplified block diagram of signal propagation. $G_1$ is the proportional derivative element.}}
   \label{fig2}
\end{figure*}

Here we consider the transfer function of the system that signal propagates in a 1-D chain with $k$ particles.
Eq.~\ref{eqdyn} shows that the acceleration of each particle (i.e., the second time derivative of particle's displacement) is affected by the position of nearby particles and itself, and the whole system is linear time-invariant.
The block diagram of this system is shown in Fig.~\ref{fig1}, where each node is the displacement of each particle $x_n$, $D$ is the delay element with transfer function $D(s)=\exp\left(-\frac{ls}{c}\right)$, $G_0$ is the proportional derivative element with transfer equation $G_0(s)=-\frac{1}{ms^2}\frac{\partial F}{\partial x}\big|_{x=l}$, $s$ is the complex variable (or complex frequency) corresponding to the Laplace transform of $f(t)$,
\begin{equation}
T_n(s)=\mathscr{L}[f(t)]=\int_0^{\infty}f(n,t)\exp(-st)\mathrm{d}t.
\end{equation}
In Fig.~\ref{fig1}, each feedback branch with a multiplicative gain of ``-2'' represents the effect of the particle position on the acceleration of itself, and the multiplicative gain of ``+'' is the effect of the posterior particle.
The feedback of the last particle ``-2'' means that the last particle is limited by the rigid boundary.
Since the element composed by a delay element and a feedback loop repeats $n$ times in the block diagram, the diagram could be simplified as shown in Fig.~\ref{fig2}, where
$$
G_1(s)=\frac{G_0D}{1+2G_0}.
$$

\begin{figure}[H]
  \centering
   \includegraphics[width=0.9\linewidth]{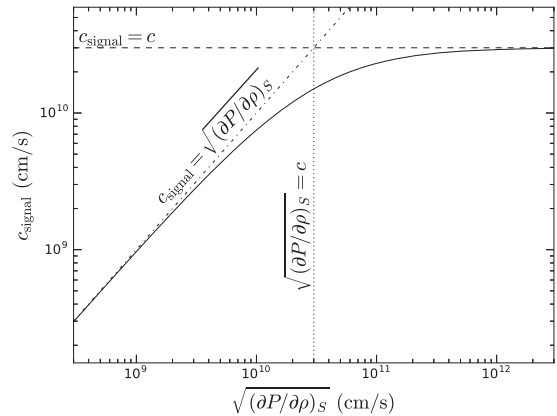}
   \caption{{\small The relation between signal speed $c_\mathrm{signal}$ and $\sqrt{\left(\frac{\partial P}{\partial\rho}\right)_S}$.}}
   \label{fig3}
\end{figure}

With Mason's gain formula, the total transfer function of the system can be obtained
\begin{equation}
G(s)=\frac{G_1^k}{\sum\limits_{m=0}^{\lfloor\frac{k}{2}\rfloor}\mathrm{C}^m_{k-m}(-G_1^2)^m}
=\frac{2^kG_1^k}{k+1+\mathcal{O}[1-4G_1^2]},
\end{equation}
With low frequency assumption $l\omega\ll\min(c,\sqrt{\left(\frac{\partial P}{\partial\rho}\right)_S})$, the transfer function can be approximated at a short-range ($1\ll k\ll\left(\frac{\partial P}{\partial\rho}\right)_S\frac{1}{l^2\omega^2}$) as following,
\begin{equation}
G(s)=\frac{2^kG_1^k}{k+1}=\frac{1}{k+1}\left[\frac{\left(\frac{\partial P}{\partial\rho}\right)_S}{l^2s^2+\left(\frac{\partial P}{\partial\rho}\right)_S}\right]^k\exp\left(-\frac{kls}{c}\right).
\end{equation}
Thus, for an impulse signal $x_0(t)=A\delta(t)$, the response should be
\begin{equation}
\begin{split}
x_k(t)&=\mathscr{L}^{-1}[\mathscr{L}[x_0(t)]G(s)]\\
&=\frac{2^{k-\frac{1}{2}}\sqrt{\pi}A}{(k+1)\Gamma(k)}\sqrt{\frac{1}{l^2}\left(\frac{\partial P}{\partial\rho}\right)_S}\theta^{k-\frac{1}{2}}J_{k-\frac{1}{2}}(\theta),
\end{split}
\label{eqresp}
\end{equation}
where $\theta=\sqrt{\frac{1}{l^2}\left(\frac{\partial P}{\partial\rho}\right)_S}\left(t-\frac{kl}{c}\right)$, $\Gamma$ is the Gamma function, and $J$ is the Bessel function of the first kind.
The position of the first maximum point of Eq.~\ref{eqresp} provides the propagating time of the signal
(here we regard the peak time as propagating time).
Considering the relation~\cite{abra70}
$$
\frac{\mathrm{d}[\theta^{\nu}J_{\nu}(\theta)]}{\mathrm{d}\theta}=\theta^{\nu}J_{\nu-1}(\theta)
$$
and the asymptotic about the first zeros of Bessel Function $j_{\nu 1}$~\cite{tric49}
$$
j_{\nu 1}=\nu+1.855757\nu^{1/3}+1.003315\nu^{-1/3}+\mathcal{O}[\nu^{-1}],
$$
the signal propagation time can be obtained,
\begin{equation}
t_{\mathrm{signal}}=\frac{k-\frac{3}{2}+1.855757\left(k-\frac{3}{2}\right)^{1/3}+\mathcal{O}[1]}
{\sqrt{\frac{1}{l^2}\left(\frac{\partial P}{\partial\rho}\right)_S}}+\frac{kl}{c}.
\label{eqspt}
\end{equation}
Finally, the signal propagating speed should take the form (for $k\gg1$),
\begin{equation}
c_{\mathrm{signal}}=\frac{kl}{t_{\mathrm{signal}}}\approx\frac{1}{\frac{1}{\sqrt{\left(\frac{\partial P}{\partial\rho}\right)_S}}+\frac{1}{c}}<c.
\label{eqcsignal}
\end{equation}
If the propagation delay of force field is ignored, i.e., $c_{\mathrm{signal}}\approx\sqrt{\left(\frac{\partial P}{\partial\rho}\right)_S}$, it is actually the conventional sound speed.
And Eq.~\ref{eqcsignal} also implicates that the signal propagating can be never faster than light,
as shown in Fig.~\ref{fig3}.

Besides, it should be noted that the speed here is just the average speed.
It is just derived from the output signal (the displacement of the last particle).
In fact, the movement of the particles except the both ends of the chain is not derived here.

\subsection{Signal propagates inside star}
\label{sect:star}

\begin{figure}[H]
  \centering
   \includegraphics[width=\linewidth]{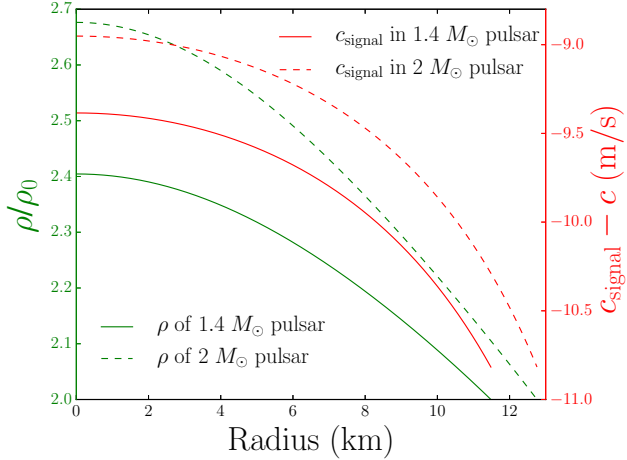}
   \caption{{\small Density $\rho$ and signal speed $c_{\mathrm{signal}}$ in the strangeon stars are shown, where $\rho_0$ is the nuclear matter density.
   The density of strangeon star is calculated with the EoS supplied by \cite{lai09}.}}
   \label{fig4}
\end{figure}

Based on the derivation in \S \ref{response}, the signal propagation speed $c_{\mathrm{signal}}$ in
strangeon matter is always smaller than speed of light.
With a definite EoS, $c_{\mathrm{signal}}$ in strangeon star
could be obtained as well.
Here, the EoS supplied by Lai \& Xu \cite{lai09} is adopted, and the corresponding $c_{\mathrm{signal}}$
in strangeon star is shown in Fig.~\ref{fig4}.
It could be evident that $c_{\mathrm{signal}}$ in strangeon star is quite close to $c$,
and $c_{\mathrm{signal}}$ decrease from stellar center to surface as same as the density.
Therefore, it would be a good approximation to assume a kinematic perturbation inside strangeon matter responses at the speed of light, $c$, because the the real one deviates only $\sim 10^{-8}c$.

\section{Discussions}
\label{sect:discussion}

\subsection{Assumptions in Derivation}

A few assumptions are adopted in the derivation, and the
applicability of these assumptions should be further discussed.

\subsubsection{small amplitude assumption}

The small amplitude assumption runs through the whole derivation process.
In fact, in traditional derivation of sound speed, small amplitude assumption is also adopted
to ensure that the system is linear, but here it is much stricter.
The small amplitude here means that the amplitude is far smaller than the interparticle spacing,
but it is always be violated in most case.
Actually, this assumption can be replaced by the stable spacing assumption, $f(n+1,t)-f(n,t)\ll [l(n+1)-l(n)]$,
i.e., the distance between nearby particles is almost invariant.
This assumption is equivalent to the traditional small amplitude assumption
(although it still doesn't apply to the normal gas).

In the derivation, we assume that the oscillation is longitudinal.
Nevertheless, the small amplitude assumption makes the derivation results
could also apply to the transverse wave.
For transverse wave, the repulsive force should be replaced by an attractive force
(the relation between this force and the tension in the chain is also different),
and the position of each particle along the chain should be arranged manually
in advance.

In the triaxial crystal, the potential in each lattice is also triaxial.
With the small amplitude assumption, the potential near each particle can be approximated
to the triaxial harmonic oscillator potential.
While the oscillation propagates along the axis of this potential (with the short
range assumption, this constraint ensures the degree of freedom to be 1), the derivations
in \S~\ref{sect:calculation} is still available.
The speed of plane sound wave along different axes is also different, depending on the potential and lattice constant.
However, the oscillation whose propagating direction avoid the potential axis is so complex that
it is not considered in this paper.

\subsubsection{low frequency assumption}

In the derivation, the frequency
is assumed low, $l\omega\ll\min(c,\sqrt{\left(\frac{\partial P}{\partial\rho}\right)_S})$, for convenience.
This assumption obviously conflicts to the input signal $x_0(t)=A\delta(t)$ in \S \ref{response}, but it
would not affect the result.
In the real physics processes with continuous time, an impulse with the form of Dirac delta function
$\delta(t)=\frac{1}{2\pi}\int_{-\infty}^{\infty}\exp(\mathrm{i}\omega t)\mathrm{d}\omega$ doesn't exist.
Real impulse can be treated as a process only contains low frequency components
$\frac{1}{2\pi}\int_{-\omega_0}^{\omega_0}\exp(\mathrm{i}\omega t)\mathrm{d}\omega$
with a very large cut-off frequency $\omega_0$ (In compact matter, $\omega_0$ can be as large as $10^{23}~\mathrm{rad/s}$, which is similar to the frequency of 100~MeV $\gamma$-ray).
This implicates that the low frequency assumption is reasonable.

In the derivation, the retarding force is adopted.
But in fact, the potential field is delayed instead of force.
The retarding potential can not only affect the action time of the force,
but also can affect the strength of the force.
With low frequency assumption, the moving speed of particles is far less than the speed of light,
and this leads to that the strength varying caused by retarding potential could be ignored.

In the longitudinal wave, the magnetic field caused by particles (if the particles are charged) moving doesn't work,
but in the transverse wave, the magnetic field could make a difference.
With low frequency assumption, this magnetic field is so weak that its effect could be ignored.
Thus, the assumption of conservative force will not lose efficacy.

Additionally, the electromagnetic radiation could be ignored.
With low frequency assumption, the variation of field energy density is too weak,
which leads us to use mass density instead of energy density in derivation.
%

\subsubsection{short-range assumption}

To simplify the equation, we adopt the ``short-range'' assumption, $k\ll\left(\frac{\partial P}{\partial\rho}\right)_S\frac{1}{l^2\omega^2}$.
But in fact, this ``short-range'' is indeed not a short distance.
For typical parameters in dense matter $\omega=10^{10}$~rad/s, $l=10^{-15}$~cm, and
$\frac{\partial P}{\partial\rho}=10^{20}~\mathrm{(cm/s)^2}$, this limit
is $10^{20}$~cm, and this value is much larger than the length scale of a compact star.

\subsubsection{rigid boundary assumption}

In \S~\ref{response}, the rigid boundary assumption is adopted to limit the
displacement of the last particle.
In fact, this boundary condition can be replaced by others, e.g.,
the free boundary condition.
For free boundary, it indeed implicates that the average point of each particle
is also the force balance point, $F(x)\big|_{x=l}=0$.
In this case, the feed back of the last particle should be adjusted to ``-''.
Nevertheless, it would not significantly change the final result of sound speed.

\subsection{Waveform Variation}

For traditional sound wave, the waveform is invariant during propagation.
But in the discrete medium, that is different.

In the particle chain, the input signal $x_0(t)$ is restricted to be continuous,
it implicates that the strength of force on other particles is continuous i.e., the second
time derivative of particle's displacement is continuous.
But the input signal not always have a continuous second time derivative.
It means that the waveform could change while propagate.
The variation of waveform indicates that the sound speed varies with frequency.
Thus, sound wave dispersion is a corollary of discrete medium.

Beside the dispersion, even though for single frequency wave, the amplitude would
also vary as shown in Eq.~\ref{eqratio}.
This effect results from the retarding force.
If the amplitude is assumed invariant, the total work of the resultant force
on each particle in one period would be nonzero.
Then, a system with sound wave with invariant amplitude is unstable,
i.e., the amplitude of sound wave must change.
The amplitude variation doesn't mean the dissipation of energy, actually the energy
is just reflected.
Just like the evanescent wave, the energy of sound wave is reflected and its amplitude decreases.

As shown in Eq.~\ref{eqresp}, the oscillation of output signal becomes
more and more violent with time, and it is obviously unreasonable.
In fact, such enhancement can not happen because of the small amplitude assumption
(although the input signal, $x_0(t)=A\delta(t)$, also violates this assumption).
If the amplitude increases to a very large value, the system would become nonlinear
and derivation in \S \ref{response} would fail.
In addition, the ringing after peak time is not overshot,
because of the amplitude of Dirac delta function is infinity (and its energy is infinity, too).
It is known that the Bessel function $J_{\nu}({\theta})$ oscillates
at a large $\theta$ with a period $2\pi$, then we can define a characteristic frequency of this system
just like eigenfrequency $\omega_\mathrm{c}=\sqrt{\frac{2}{l^2}\left(\frac{\partial P}{\partial\rho}\right)_S}$.
This frequency describes the oscillation property of a chain, and it is also the limit of the ``low frequency''.

\subsection{Sound Speed and Strangeon Star}

With the traditional sound speed formula, it has been proved that the
mass of a neutron star should be less than $3.2~M_{\odot}$ \cite{rhoa74}, and
a ``safe'' upper limit for neutron star mass of $2.9~M_{\odot}$ can
also be obtained \cite{kalo96}.
But in strangeon matter, the sound speed is no more a limit of the matter state.
Then, the maximum mass of strangeon star could easily exceed $3.2~M_{\odot}$,
and it can even be much higher.

This result may explain the ``mass gap'' puzzle.
It is found that there could be ``gap'' between the least massive black hole and
the upper limit for neutron star masses~\cite{bail98}.
The lower bound of the 1\% quantile from each black hole
mass distribution is also shown as about $4.3~M_{\odot}$~\cite{farr11}.
With the strangeon star model, this puzzle could be explained if the maximum mass
of strangeon star can reach $\sim4~M_{\odot}$.

Besides, having a value of $\partial P/\partial\rho$ which can be larger than $c^2$ will also result
in a quite different tidal deformability of a strangeon star.
For conventional nuclear EoSs, it's widely accepted that the asymptotic sound speed is
smaller than $c/\sqrt{3}$ for ultra high density.
By fixing an upper limit for $\partial P/\partial\rho$ accordingly, it has been shown
that the maximum mass of an EoS model decreases as the tidal deformability decreases~\cite{anna17}.
Similar arguments is found for conventional quark stars within the MIT bag model description and
a consideration of color-flavor-locked phase~\cite{zhou17}.
Now that the strangeon star can have a value of $\partial P/\partial\rho$ larger than $c^2$ without
violating causality, it is found that the tidal deformability of it is compatible with the
observation of GW170817~\cite{abbo17} although its maximum mass is very large~\cite{lai17}.

\section{Summary}
\label{sect:summary}

We calculate the oscillation propagation in discrete medium in frequency
domain and time domain respectively, and obtain the sound speed.
Our results show that the signal propagation speed would never exceed the speed of light,
and in small $(\partial P/\partial\rho)_S$ case its expression would degenerate
to tradition form of sound speed.
Thus, the strangeon star model can be safely used without worrying about
if it conflicts to causality.

In the strangeon star model, the mass of pulsars could be much higher.
It implies that more massive pulsars could be found, but the accurate mass
of pulsar is hard to measure.
In the future, FAST (Five-hundred-meter Aperture Spherical radio Telescope)
could be sensitive to detect more weaker radio signal from pulsars far away,
and it could also provide higher precision timing result to obtain the accurate
mass of pulsars \cite{lu16}.
More massive pulsars are expected.

\Acknowledgements{This work is supported by the National Key R\&D Program of China (No. 2017YFA0402600), the National Natural Science Foundation
of China (Grant No. 11225314), and the Open Project Program of the Key Laboratory of Radio Astronomy, Chinese Academy of Sciences.
The FAST FELLOWSHIP is supported by Special Funding for Advanced Users, budgeted and
administrated by Center for Astronomical Mega-Science, Chinese Academy of Sciences (CAMS).
I would like to thank my friend Chuzhong Pan
for his advice.}




\end{multicols}

\begin{thebibliography}{99}

\bibitem{abbo17}
B.~P. Abbott, R. Abbott, T.~D. Abbott,, et al., Physical Review Letters, 119, 161101 (2017).

\bibitem{baio17}
L. Baiotti, \& L. Rezzolla, Reports on Progress in Physics, 80, 096901 (2017).

\bibitem{rezz13}
L. Rezzolla \& O. Zanotti, 2013, Relativistic Hydrodynamics, by L.~Rezzolla and O.~Zanotti.~Oxford University Press, 2013.~ISBN-10: 0198528906; ISBN-13: 978-0198528906.

\bibitem{bodm71}
A.~R. Bodmer, \prd, 4, 1601 (1971).

\bibitem{witt84}
E. Witten, \prd, 30, 272 (1984).

\bibitem{alco86}
C. Alcock, E. Farhi \& A. Olinto, \apj, 310, 261 (1996).

\bibitem{haen86}
P. Haensel, J.~L. Zdunik \& R. Schaefer, \aap, 160, 121 (1986).

\bibitem{demo10}
P.~B. Demorest, T. Pennucci, S.~M. Ransom, M.~S.~E. Roberts \& J.~W.~T. Hessels, \nat, 467, 1081 (2010).

\bibitem{anto13}
J. Antoniadis, P.~C.~C. Freire, N. Wex, et al., Science, 340, 448 (2013).

\bibitem{scha08}
J. Schaffner-Bielich, Nuclear Physics A, 804, 309 (2008).

\bibitem{Bombaci17}
I. Bombaci, JPS Conf. Proc., {\bf 17}, 101002 (2017)

\bibitem{xu03}
R.~X. Xu, \apj, 596, L59 (2003).

\bibitem{lai09}
X.~Y. Lai, \& R.~X. Xu, \mnras, 398, L31 (2009).

\bibitem{qiao04}
G.~J. Qiao, K.~J. Lee, B. Zhang, R.~X. Xu, \& H.~G. Wang, \apjl, 616, L127 (2004).

\bibitem{lu13}
J.~G. Lu, R.~X. Xu, \& H. Feng, Chinese Physics Letters, 30, 059501 (2013).

\bibitem{zhou14}
E.~P. Zhou, J.~G. Lu, H. Tong \& R.~X. Xu, \mnras, 443, 2705 (2014).

\bibitem{wang17}
W.~Y. Wang, J.~G. Lu, H. Tong, et al., \apj, 837, 81 (2017).

\bibitem{bayi80}
S.~{\c S}. Bayin, \prd, 21, 1503 (1980).

\bibitem{blud68}
S.~A. Bludman, \& M.~A. Ruderman, Physical Review, 170, 1176 (1968).

\bibitem{blud70}
S.~A. Bludman, \& M.~A. Ruderman, \prd, 1, 3243 (1970).

\bibitem{keis96}
B.~D. Keister \& W.~N. Polyzou, \prc, 54, 2023 (1996).

\bibitem{rhoa74}
C.~E. Rhoades \& R. Ruffini, Physical Review Letters, 32, 324 (1974).

\bibitem{guo14}
Y.~J. Guo, X.~Y. Lai \& R.~X. Xu, Chinese Physics C, 38, 055101 (2014).

\bibitem{lai13}
X.~Y. Lai, C.~Y. Gao, \& R.~X. Xu, \mnras, 431, 3282 (2013).

\bibitem{capo79}
G. Caporaso, \& K. Brecher, \prd, 20, 1823 (1979).

\bibitem{lai17}
X.~Y. Lai, Y.~W. Yu, E.~P. Zhou, Y.~Y. Li, and R.~X. Xu,\ RAA, {\bf 18}, 24 (2018).

\bibitem{abra70}
M. Abramowitz, \& I. A. Stegun,, (ed.) Handbook of Mathematical Functions:
With Formulas, Graphs, and Mathematical Tables, National Bureau of Standards, 1970.

\bibitem{tric49}
F.~G. Tricomi, Atti Accad. Sci. Torino Cl. Fis. Mat. Nature., 83, 3 (1949).

\bibitem{kalo96}
V. Kalogera \& G. Baym, \apj, 470, L61 (1996).

\bibitem{bail98}
C.~D. Bailyn, R.~K. Jain, P. Coppi, \& J.~A. Orosz, \apj, 499, 367 (1988).

\bibitem{farr11}
W.~M. Farr, N. Sravan, A. Cantrell, et al., \apj, 741, 103 (2011).

\bibitem{anna17}
E. Annala, T. Gorda, A. Kurkela, \& A. Vuorinen, arXiv:1711.02644 (2017).

\bibitem{zhou17}
E.~P. Zhou, X. Zhou, A. Li, arXiv:1711.04312 (2018).

\bibitem{lu16}
J. Lu, \& R. Xu, Frontiers in Radio Astronomy and FAST Early Sciences Symposium 2015, 502, 35 (2016).

\end{thebibliography}
\end{document}